\documentclass[aps,prd,twocolumn,twoside,floatfix,preprintnumbers,nofootinbib]{revtex4}

\usepackage{graphicx}
\usepackage{amsmath}
\usepackage{enumerate}

\begin{document}
\preprint{MPP-2010-109}

\title{Phenomenology of SUSY SU(5) with type I+III seesaw}

\author{Carla~Biggio}\email{biggio@mppmu.mpg.de}
\author{Lorenzo~Calibbi}\email{calibbi@mppmu.mpg.de}

\affiliation{Max-Planck-Institut f\"ur Physik (Werner-Heisenberg-Institut), D-80805 M\"unchen, Germany}

\begin{abstract}

We consider a supersymmetric SU(5) model where two neutrino masses are obtained via a mixed type I+III seesaw mechanism induced by the component fields of a single SU(5) adjoint. We have analyzed the phenomenology of the model paying particular attention to flavour violating processes and dark matter relic density, assuming universal boundary conditions. We have found that, for a seesaw scale larger than $10^{12\div 13}$~GeV, BR$(\mu\to e \gamma)$ is in the reach of the MEG experiment in sizable regions of the parameter space. 
On the other side, current bounds on it force BR$(\tau \to \mu \gamma)$ to be well below the reach of forthcoming experiments, rendering thus the model disprovable if a positive signal is found. The same bounds still allow for a sizable positive contribution to $\epsilon_K$, while the CP violation in the $B_s$ mixing turns out to be too small to account for the di-muon anomaly reported by the D0 collaboration. 
Finally, the regions where the neutralino relic density is within the WMAP bounds can be strongly modified with respect to the
constrained MSSM case. In particular, a peculiar coannihilation region, bounded from above, can be realized, which 
allows us to put an upper bound on the dark matter mass for certain set-ups of the parameters. 

\end{abstract}

\maketitle

\section{Introduction}\label{sec:intro}

Neutrino masses are an indication for the presence of new physics beyond the standard model (SM). The simplest extension consisting in adding to the SM fields three right-handed (RH) neutrinos and giving them a Dirac mass is not very satisfactory since it would require extremely small Yukawa couplings, much smaller then the ones for charged particles. A nice way of explaining neutrino masses as well as their smallness is through the so-called seesaw mechanism: new fields with masses much heavier than the electroweak (EW) scale, once integrated out, generate a dimension-five operator which, after EW symmetry breaking, gives a Majorana mass to neutrinos. When neutrino masses are generated at tree-level, three heavy mediators are possible, namely singlet fermions (corresponding to the so-called type-I seesaw~\cite{typeI}), triplet scalar (type II~\cite{typeII}) and triplet fermions (type III~\cite{Foot:1988aq})\footnote{See Ref.~\cite{Abada:2007ux} for a review on the three mechanisms.}. Independently of the nature of the mediators, the neutrino mass results to be suppressed by their heavy mass. Interestingly, $\mathcal{O}(1)$ Yukawa couplings require the scale of new physics to be around the grand unification scale. It is then natural to study the seesaw mechanisms in the context of grand unified theories (GUT). Moreover, as it is well known, the presence of low scale supersymmetry (SUSY) triggers the unification of gauge couplings, so that usually SUSY GUTs are considered. 

In the literature plenty of models of SUSY GUTs including a seesaw mechanism for neutrino mass generation has been proposed. Actually, many different GUTs have been studied, involving the different seesaws, especially the types I and II. As for the type III (and type I+III), its embedding in a SUSY GUT was firstly proposed in Ref.~\cite{Ma:1998dn} and then discussed in Ref.~\cite{Perez:2007iw} (in the context of a renormalizable model) and in Ref.~\cite{FileviezPerez:2009im}, and embedded in a flavour model in Ref.~\cite{Cooper:2010ik}. In all these cases the grand unified group considered is SU(5), which is the simplest group in which the SM gauge group can be embedded 
(for a wide discussion about SUSY SU(5) with the three different seesaw mechanisms, see \cite{Borzumati:2009hu}).
In a minimal version of SU(5)~\cite{Georgi:1974sy} neutrinos are massless, so that this GUT model has to be extended in order to account for neutrino masses. The model we consider here is somehow the simplest extension of a SUSY SU(5) accounting for neutrino masses, since with the simple addition of only one SU(5) representation two neutrino masses are generated via a mixed type I+III seesaw mechanism~\cite{FileviezPerez:2009im}. This is different from other models, where at least two representations were needed to account for two or three neutrino masses. Indeed, in the case of type I seesaw, the addition of at least two singlets is mandatory, while for the type II one must include both a triplet and its conjugate (or, in terms of SU(5) multiplets, a {\bf 15} and a $\bf \bar{15}$~\cite{Rossi:2002zb})\footnote{Also in the type III models mentioned before more representations are introduced, either matter~\cite{Cooper:2010ik} or Higgs~\cite{Perez:2007iw}.}. This model is the SUSY version of the model firstly proposed in Ref.~\cite{Bajc:2006ia, Bajc:2007zf}. Also there two neutrino masses are obtained via a mixed type I+III seesaw, but being the model non-SUSY, the spectrum is completely different. Indeed, to guarantee unification, the triplet must be at the TeV scale, while the singlet mass is not specified. In this model lepton flavour violation (LFV) is usually suppressed either by the small Yukawas or by a large mass. This is the standard situation in non-SUSY seesaw models, unless a low-scale inverse seesaw is realized \cite{Abada:2007ux,Abada:2008ea}. Furthermore, in presence of cancellations in the neutrino mass matrix, sizable rates of LFV processes are still possible (see for example, in the context of type I+III seesaw, \cite{Kamenik:2009cb}). 

Here we study in detail the phenomenology of this ``minimal'' SUSY SU(5) with massive neutrinos, paying particular attention to the effects on LFV processes. In order to isolate the effects purely induced by the RG running of the SUSY parameters we assume universal soft masses at the GUT scale. We also discuss possible related effects in the quark sector and how the region of the parameter space where the lightest SUSY particle (LSP) is a viable dark matter (DM) candidate are modified. The rest of the paper is organized as follows: in Sect.~\ref{Sec:2} we introduce the model, in Sect.~\ref{Sec:3} we discuss the flavour violation, in Sect.~\ref{Sec:4} we present numerical results and in Sect.~\ref{Sec:5} we conclude. The renormalization group equations (RGEs) of this model are gathered in the Appendix.


\section{The model}\label{Sec:2}

We consider a SUSY SU(5) model where the matter content is enlarged
with a {\bf 24} representation of SU(5), in order to get neutrino
masses. Indeed the {\bf 24} can be decomposed under ${\rm SU(3)}_c\times{\rm SU(2)}_L\times{U(1)}_Y$ 
as $S+T+O+V+\bar{V}$, where
\begin{equation*}
\label{eq:24dec}
S\sim(1,1)_0 \quad T\sim(1,3)_0 \quad O\sim(8,1)_0 \quad V\sim(3,2)_{-5/6} .
\end{equation*}
$S$ and $T$ give mass to two neutrinos via a mixed type I+III seesaw
mechanism. The relevant superpotential terms are then
\begin{eqnarray}
\label{eq:WSU(5)}
W_{SU(5)}&=&\frac{1}{4}(Y_{10})_{ij} \ {\bf 10}_i  {\bf 10}_j  {\bf 5}_H  
+ \sqrt{2} (Y_{\overline 5})_{ij} \ {\bf 10}_i  \overline{\bf 5}_j 
 \overline{\bf 5}_H +\nonumber \\ 
&& y_{24}^{i} \ \overline{\bf 5}_i  {\bf 24}\, {\bf 5}_H 
+\frac{M_{24}}{2} {\bf 24}\, {\bf 24}
+ \cdots
\end{eqnarray}
Here ${\bf \bar{5}}=(d^c,L)$, ${\bf 10}=(u^c,Q,e^c)$ and ${\bf
\bar{5}_H}=(H_T,H_1)$, as usual. 
Below the SU(5) breaking scale, the superpotential reads:
\begin{eqnarray}
\label{eq:WSU(5)broken}
W &=&  (Y_u)_{ij}\, u^c_i Q_j h_u + (Y_d)_{ij}\, d^c_i Q_j h_d
+ (Y_e)_{ij}\, e^c_i L_j h_d +  \nonumber  \\
&& y_T^i\, L_i T h_u
+ y_S^i\, L_i S h_u  
+  y_V^i\, d^c_i V h_u  +  \nonumber  \\
&& \frac{M_s}{2} S S +\frac{M_T}{2} T T +\frac{M_O}{2} O O+ M_V \, V \overline{V} .
\end{eqnarray}
Notice that, from Eq.~(\ref{eq:WSU(5)}), it follows that the $SU(3)$ octet $O$ 
and the field $\overline{V}$
do not have Yukawa interactions with fields which remain lighter than the GUT scale.

From Eq.~(\ref{eq:WSU(5)broken}), it is easy to see that the
singlet $S$ and the neutral component of the triplet $T$ generate neutrino
masses via a seesaw mechanism:
\begin{equation}
\label{eq:numass}
m_\nu=-\frac{v_u^2}{2}\left(\frac{y_S^i y_S^j}{M_S}
                          +\frac{y_T^i y_T^j}{M_T}\right).
\end{equation}
Notice that, since $S$ and $T$ belong to the same SU(5) multiplet,
at the GUT scale $\vec{y}_T= \sqrt{10/3}\ \vec{y}_S= \vec{y}_{24}/\sqrt{2}$. 
The previous $m_\nu$ is then a
rank-1 matrix and only one neutrino mass is generated. In principle
the Yukawas could be misaligned in the running from the GUT scale to
the seesaw scale simply via RGE effects, but in practice
the resulting misalignment is too small, giving rise to a second
neutrino mass much smaller that the solar mass. It is then clear that
the GUT relation on the Yukawa couplings must be somehow altered, in
order to get two massive neutrinos. Since non-renormalizable operators
should anyway be added to the SUSY SU(5) lagrangian in order to break
the unwanted relation $Y_e=Y_d^T$ which is in disagreement with the
experimental value of fermion masses~\cite{Ellis:1979fg}\footnote{An
  alternative solution is adding new Higgs
  representation~\cite{Georgi:1979df}.}, 
we can also add non-renormalizable operators involving the {\bf 24} \cite{Bajc:2006ia}:
\begin{eqnarray}
\label{eq:nonrinyukawas}
W_{\rm NR} &\supset&
\frac{1}{\Lambda} \left[y_1^i\, ({\bf\bar{5}}_i \,{\bf 24}) ({\bf 24_H}\,{\bf 5_H})
   + y_2^i\, ({\bf\bar{5}}_i \,{\bf 24_H}) ({\bf 24}\,{\bf 5_H}) +\right.\nonumber \\
&& \left. y_3^i\, {\bf\bar{5}}_i {\rm tr}({\bf 24}\,{\bf 24_H}){\bf 5_H}\right] \,,
\end{eqnarray}
where $\Lambda$ represents the cut-off of the theory (e.g. the Planck scale $M_{\rm Pl}$)
and ${\bf 24_H}$ is the Higgs in the adjoint representation which breaks SU(5), with
$\langle{\bf 24_H}\rangle = diag(1/3,1/3,1/3,-1/2,-1/2)\sqrt{3/5}\,v_{\rm G}$. The three above operators could be for instance generated by integrating-out, respectively: a Higgs superfield in the ${\bf 45}$+${\bf\bar{45}}$ representation, a matter ${\bf 45}$+${\bf \bar{45}}$, a SU(5) singlet, all having a mass of the order of the scale $\Lambda$.

After SU(5) breaking, the Yukawa couplings of the different components
of the {\bf 24} split and the matching at the GUT scale reads:
\begin{eqnarray}
\label{eq:yT}
\vec{y}_T&=& \frac{1}{\sqrt{2}}\vec{y}_{24}+ \frac{1}{2}\sqrt{\frac{3}{10}}\frac{v_{\rm G}}{\Lambda}(\vec{y}_1+\vec{y}_2),\\
\label{eq:yS}
\vec{y}_S&=& \sqrt{\frac{3}{20}}\vec{y}_{24}+
\frac{3}{20}\frac{v_{\rm G}}{\Lambda}(\vec{y}_1+\vec{y}_2) + 
\frac{1}{2}\frac{v_{\rm G}}{\Lambda}\vec{y}_3 ,\\
\label{eq:yV}
\vec{y}_V&=& \frac{1}{\sqrt{2}}\vec{y}_{24}+\frac{1}{2}\sqrt{\frac{3}{10}}\frac{v_{\rm G}}{\Lambda}\vec{y}_1-
        \frac{1}{2}\sqrt{\frac{2}{15}}\frac{v_{\rm G}}{\Lambda}\vec{y}_2 .
\end{eqnarray}
In practice the coupling $\vec{y}_3$ determines the misalignment between
$\vec{y}_S$ and $\vec{y}_T$ that permits to generate two non-zero neutrino
masses. As we will see later, LFV processes suggest for the seesaw scale an upper bound around $10^{13\div 14}$~GeV, corresponding 
to neutrino Yukawa couplings smaller than $10^{-(1\div 2)}$. From Eq.~(\ref{eq:yS}) we then deduce that $\vec{y}_3$ should be $\mathcal{O}$(1) and the cutoff scale $\Lambda$ cannot be too large, if we want to saturate these bounds. 
From now on we will consider $\vec{y}_S$ and $\vec{y}_T$ as independent parameters. 

In the same way as the splitting among
the Yuwakas is generated via SU(5) breaking effect, also a splitting
in the masses arise. 
The terms contributing to the masses of the {\bf 24} fields are
\begin{eqnarray}
\label{eq:nonrinmasses}
\frac{1}{2}M_{24}{\rm tr}\left({\bf 24}^2\right)+
\lambda_{24}{\rm tr}\left({\bf 24}^2\, {\bf 24_H}\right) 
+ \mathcal{O}\left(\frac{v_{\rm G}^2}{\Lambda}\right)\, ,
\end{eqnarray}
giving rise to the following masses:
\begin{eqnarray}
\label{eq:mS}
M_S &=& \frac{M_{24}}{2}-\lambda_{24}\frac{v_{\rm G}}{2\sqrt{60}}+
\lambda^{\rm NR}_S\frac{v_{\rm G}^2}{\Lambda},\\
\label{eq:mT}
M_T &=& \frac{M_{24}}{2}-\lambda_{24}\frac{3v_{\rm G}}{2\sqrt{60}}+
\lambda^{\rm NR}_T\frac{v_{\rm G}^2}{\Lambda},\\
\label{eq:mO}
M_O &=& \frac{M_{24}}{2}+\lambda_{24}\frac{2v_{\rm G}}{2\sqrt{60}}+
\lambda^{\rm NR}_O\frac{v_{\rm G}^2}{\Lambda},\\
\label{eq:mV}
M_V&=&\frac{M_{24}}{2}-\lambda_{24}\frac{v_{\rm G}}{4\sqrt{60}}+
\lambda^{\rm NR}_V\frac{v_{\rm G}^2}{\Lambda} , 
\end{eqnarray}
where the actual expressions of the $\lambda^{\rm NR}_X$ in terms
of the possible NR operators couplings can be found in \cite{Bajc:2006ia}. 

If we do not allow fine cancellations, we can conclude that the components 
of the {\bf 24} have masses of the same order of magnitude, so that only small threshold effects are introduced and the successful
gauge coupling unification of the MSSM is maintained. 
Notice that, since LFV processes disfavor a seesaw scale larger than $10^{14}$~GeV, the coupling $\lambda_{24}$ should be $10^{-(1\div 2)}$ or smaller while the non-renormalizable couplings $\lambda^{\textrm{NR}}_i$ can be larger, depending on the cutoff scale $\Lambda$. 

We now focus on neutrino masses. Eq.~(\ref{eq:numass}) can be
inverted and the Yukawa couplings can be expressed in terms of low
energy parameters. Since in this case we have only two massive
neutrinos, Casas-Ibarra parameterization is simplified, since the R
matrix depends now only on one complex parameter $z$~\cite{Ibarra:2003up}. The Yukawa couplings can then be expressed as \cite{Bajc:2007zf}:
\begin{equation}
\label{eq:tn}
vy_T^{i*}=
 \left\{ \begin{array} {cc}
 i\sqrt{M_T}\left( U_{i2}\sqrt{m^\nu_2}\cos{z}+U_{i3}\sqrt{m^\nu_3}\sin{z}\right)
   &  {\rm NH} \\
i\sqrt{M_T}\left(U_{i1}\sqrt{m^\nu_1}\cos{z}+U_{i2}\sqrt{m^\nu_2}\sin{z}\right) 
   &  {\rm IH} 
\end{array} \right. 
\end{equation}
and
\begin{equation}
\label{eq:tns}
vy_S^{i*}=
 \left\{ \begin{array} {cc}
-i \sqrt{M_S}\left( U_{i2}\sqrt{m^\nu_2}\sin{z}-U_{i3}\sqrt{m^\nu_3}\cos{z}\right)
   &  {\rm NH} \\
-i \sqrt{M_S}\left(U_{i1}\sqrt{m^\nu_1}\sin{z}-U_{i2}\sqrt{m^\nu_2}\cos{z}\right)
   &  {\rm IH} 
\end{array} \right. 
\end{equation}
There is another solution with the opposite sign for the second terms
in Eqs.~(\ref{eq:tn},~\ref{eq:tns}). 
In the above equations, $U_{ij}$ are elements of the PMNS matrix\footnote{We remind the reader that here, 
like in the 2 RH neutrinos case, the PMNS matrix only has two phases: a Dirac phase $\delta$ and a Majorana phase $\Phi$.}.  
For the neutrino mass eigenvalues we have in the normal hierarchy (NH) case: 
\begin{equation}
m_1^\nu=0 \quad m_2^\nu=\sqrt{\Delta m_S^2}  \quad m_3^\nu=
\sqrt{\Delta m_A^2+\Delta m_S^2}  ,
\end{equation}
while in the inverted hierarchy (IH) case neutrino masses are given by:
\begin{equation}
m_1^\nu=\sqrt{\Delta m_A^2-\Delta m_S^2}  \quad
m_2^\nu=\sqrt{\Delta m_A^2} \quad m_3^\nu=0 ,
\end{equation}
where we take the neutrino mass parameters \cite{GonzalezGarcia:2010er}
as measured in the solar and atmospheric oscillation experiments
\begin{eqnarray}
\label{eq:expmasses}
&&\Delta m_S^2 \approx 7.59 \times 10^{-5}\ {\rm eV}^2,\\ 
&&\Delta m_A^2 \approx 2.46\ (2.36) \times 10^{-3}\ {\rm eV}^2 
\quad {\rm NH}\ ({\rm IH}).
\end{eqnarray}
From Eqs.~(\ref{eq:tn},~\ref{eq:tns}) we see that the Yukawa
couplings grow with the square root of the mass of the heavy fermions,
as a trivial consequence of the seesaw formula Eq.~(\ref{eq:numass}),
and increase exponentially when Im($z$) grows. If we want to avoid
unnatural cancellations in the neutrino sector (e.g. between the two terms of Eq.~(\ref{eq:numass})),
Im($z$) should be $\le \mathcal{O}(1)$. As we will show later, the
present bound on $\mu\to e\gamma$ can actually constrain it to smaller
values.

Notice that in what respect neutrino masses, this model is not
different from a model with two right-handed neutrinos (2RHN) \cite{Ibarra:2003up, Ibarra:2005qi, Guo:2006qa}, 
where for instance the same parameterization of Eqs.~(\ref{eq:tn},~\ref{eq:tns})
holds. However, this model, besides the fact of being better motivated 
from a GUT perspective, has got some features which distinguish it from 
a generic 2RHN model:
\begin{itemize}
 \item Up to fine-tuning, the parameter space is
more restricted, since it is natural to assume $M_S$ and $M_T$ to be
of the same order of magnitude (see Eqs.~(\ref{eq:mS},~\ref{eq:mT})). 
Moreover, barring cancellations, $y^i_S$ and $y^i_T$ 
will be also of the same order of magnitude (Eqs.~(\ref{eq:yT},~\ref{eq:yS})).
 \item The presence of the SU(5) partners of $S$ and $T$
induces flavour violating effects in the hadronic sector as well, 
similarly to what happens in the leptonic sector. Again, the relevant
couplings $y^i_V$, even if in general independent, are expected to be
of the same order of $y^i_S$ and $y^i_T$. 
\item The presence of a full {\bf 24} at an intermediate scale 
between the GUT and the EW scales does not spoil gauge coupling unification
if $M_T \simeq M_O \simeq M_V \simeq M_I$ , as in our case, but affects 
the gauge couplings running above $M_I$. This can have an impact on the 
SUSY spectrum and, in particular, on the regions of the parameter space 
which provide a relic density for the LSP within the WMAP bounds, as we are
going to discuss in section \ref{Sec:4}. 
\end{itemize}

\section{Flavour violating processes}\label{Sec:3}

The presence of the fields in the ${\bf 24}$ modifies the renormalization group running of the parameters of the model, 
both the superpotential couplings and the SUSY breaking terms. 
For instance, the renormalization group equations (RGEs) for the scalar masses are now given by:
\begin{equation}
\label{eq:rge}
16\pi^2\frac{d}{dt} m_\phi^2=\beta^{\rm MSSM, 1}_{m_\phi^2} + \beta^{{\bf 24},1}_{m_\phi^2} ,
\end{equation}
where $\beta^{\rm MSSM,1}_{m^2_\phi}$ is the usual MSSM 1-loop $\beta$-function and
$\beta^{{\bf 24},1}_{m^2_\phi}$ is the new 1-loop contribution given by the
new fields in the {\bf 24}, with clearly $\beta^{{\bf 24},1}_{m_\phi^2}\neq 0$ only above the {\bf 24}
energy scale.

In particular, the couplings of the seesaw fields, $S$ and $T$, with the lepton doublet will affect
the running of the left-handed (LH) slepton masses, generating off-diagonal flavour violating entries, 
in perfect analogy with what happens in the context of supersymmetric
seesaw of type I \cite{borzumati-masiero, Ibarra:2003up} and type II \cite{Rossi:2002zb}. 
In addition, the presence of the SU(5) partner, $V$, of the seesaw fields 
will induce an analogous effect for the RH down squarks.

The complete RGEs of the model are given in the Appendix.
Let us display here the $\beta$-functions of, respectively, the LH slepton and RH down-squark soft masses,
which are the relevant ones for outlining the effects mentioned above:
\begin{eqnarray}
\label{eq:betamL}
\left(\beta^{{\bf 24},1}_{m_{\tilde L}^2}\right)_{ij}\!\!&=&\!
\frac{3}{2}\Big(
y^*_{Ti}\, (y_T^T\, m_{\tilde L}^2)_j\, +\, (m_{\tilde L}^2\, y_T^*)_i\, y^T_{Tj}+\nonumber\\
\!\!&&\! 2\, y^*_{Ti}\,y^T_{Tj}\,(m_{H_u}^2+m_{\tilde T}^2)+2\, A^*_{Ti}\,A^T_{Tj}\Big)+\nonumber\\
\!\!&&\! \Big(
y^*_{Si}\, (y_S^T\, m_{\tilde L}^2)_j\, +\, (m_{\tilde L}^2\, y_S^*)_i\, y^T_{Sj}+\nonumber\\
\!\!&&\! 2\, y^*_{Si}\,y^T_{Sj}\,(m_{H_u}^2+m_{\tilde S}^2)+2\, A^*_{Si}\,A^T_{Sj}\Big),
\\
\label{eq:betamd}
\left(\beta^{{\bf 24},1}_{m_{\tilde{d}^c}^2}\right)_{ij}\!\!&=&\!
2\Big(
y^*_{Vi}\, (y_V^T\, m_{\tilde{d}^c}^2)_j\, +\, (m_{\tilde{d}^c}^2\, y_V^*)_i\, y^T_{Vj}+\nonumber\\
\!\!&&\!2\, y^*_{Vi}\,y^T_{Vj}\,(m_{H_u}^2+m_{\tilde S}^2)+2\, A^*_{Vi}\,A^T_{Vj}\Big).
\end{eqnarray}
Off-diagonal flavour violating entries in the LH slepton and RH down-squark mass matrices are then 
generated by RG running from $M_{\rm GUT}$ down to the seesaw fields mass scales, even starting 
with universal boundary conditions at $M_{\rm GUT}$, $m_{{\tilde L}\,,\tilde{d}^c}^2 = m^2_0 \mathbf{1}$.
From Eqs.~(\ref{eq:betamL},~\ref{eq:betamd}), we can estimate
the flavour violating mass-insertions, $\delta_{i\neq j}\equiv m^2_{ij}/\sqrt{m^2_{ii}m^2_{jj}}$,
which parameterize the amount of flavour violation induced by the running. At leading-log, they read:
\begin{align}
\label{eq:deltaLLe}
(\delta^e_{\rm LL})_{ij} &=\frac{1}{8\pi^2}\frac{(3 m^2_0+A_0^2)}{\overline{m}^2_{\tilde{L}}} \times\nonumber \\
 &\left[\frac{3}{2}y_T^{i*}y_T^j \ln\left( \frac{M_{\rm GUT}}{M_T}\right)+
       y_S^{i*}y_S^j \ln\left(\frac{M_{\rm GUT}}{M_S}\right)\right], \\
\label{eq:deltaRRd}
(\delta^d_{\rm RR})_{ij} &=\frac{1}{8\pi^2}\frac{(3 m^2_0+A_0^2)}{\overline{m}^2_{\tilde{d}^c}}\
2\,y_V^{i*}y_V^j \ln\left(\frac{M_{\rm GUT}}{M_V}\right),
\end{align}
where $\overline{m}^2_{\tilde{L}}$, $\overline{m}^2_{\tilde{d}^c}$ are average slepton and squark squared masses
at low energy.
Eqs.~(\ref{eq:deltaLLe},~\ref{eq:deltaRRd}) provide a good estimate of the FV mass-insertions, unless $m_0$ 
is too small. In the case of $m_0 \simeq 0$, which is indeed possible in the model as we will discuss in the next
section, the sfermion masses are generated by the running, but Eqs.~(\ref{eq:deltaLLe},~\ref{eq:deltaRRd})
are clearly not valid anymore, since sfermion masses are vanishing at $M_{\rm GUT}$ and possible
off-diagonal entries in the mass matrices can be only generated at orders higher than the leading-log.

Keeping that in mind, we can still make use of Eq.~(\ref{eq:deltaLLe}) to get an idea of the expected
amount of LFV. For instance, in the case $M_S \simeq M_T \simeq 10^{13}$ GeV, we see from the seesaw formula, 
Eq.~(\ref{eq:numass}), that typically $y_{S,T}^i \simeq \mathcal{O}(10^{-2})$ and therefore\,
assuming $\overline{m}^2_{\tilde{L}} \simeq m_0^2$ and $A_0 \simeq 0$, Eq.~(\ref{eq:deltaLLe}) gives roughly:
\begin{equation}
\label{eq:dLLestimate}
(\delta^e_{\rm LL})_{ij} \simeq \mathcal{O}(10^{-4}) \quad \textrm{for}\ M_{S,T} \simeq 10^{13}~{\rm GeV}\,,  
\end{equation}
value which can give sizable effects in the $\mu$-$e$ transitions only and can already exclude 
the SUSY parameter space in the light sleptons regime\footnote{Cfr. the bounds provided in \cite{lfv-bounds}.}.

We can also estimate the typical ratio of the BRs of different LFV processes. 
Given that the main source of LFV is represented by the $(\delta^e_{\rm LL})_{ij}$, we have:
\begin{equation}
\frac{{\rm BR}(\ell_i\to \ell_j\gamma)}{ {\rm BR}(\ell_i\to \ell_j \nu\bar{\nu})}\propto|(\delta^e_{\rm LL})_{ij}|^2\,,
\end{equation}
hence the ratio of BRs in the $\tau$-$\mu$ and $\mu$-$e$ channels can be estimated to be:
\begin{equation}
\label{eq:R}
R\equiv\frac{{\rm BR}(\tau\to \mu\gamma)}{{\rm BR}(\mu\to e\gamma)}
\simeq 0.17\times
\frac{\left|y_S^{3*}y_S^2+\frac{3}{2}\, y_T^{3*}y_T^2\right|^2}
{\left|y_S^{1*}y_S^2+\frac{3}{2}\, y_T^{1*}y_T^2\right|^2}.
\end{equation}
Using Eqs.~(\ref{eq:tn},~\ref{eq:tns}), one can check that $4\lesssim R \lesssim 80$ in the normal hierarchy case
for a real parameter $z$, if $U_{e3}\simeq 0$ and the Majorana phase $\Phi$ vanishes as well. As the value of $U_{e3}$ increases, one can verify that $R$ diminishes and it becomes $\mathcal{O}(1)$ for $U_{e3}\simeq 0.2$. 
Interestingly, as soon as Im($z$) is switched on, $R$ rapidly drops to $\mathcal{O}(1)$ values independently of the value of $U_{e3}$. As for the role of the phases, they also generically tend to reduce $R$, even if for small non-zero values of $U_{e3}$ the Dirac phase somehow compensates the $U_{e3}$ effect, preventing the reduction of the ratio. Moreover, the presence of the phases increases $(\delta^e_{\rm LL})_{12}$, from which a bound on Im($z$) can be derived (see later).
In the inverted hierarchy case with a real $z$, $R$ can even diverge, since for certain values of $z$ $(\delta^e_{\rm LL})_{12}$
can vanish. However, such cases correspond to set-up of the Yukawas (e.g. $y_{S, T}^i \ll y_{T, S}^i$) which cannot be considered
natural in the light of Eqs.~(\ref{eq:yT},~\ref{eq:yS}). Moreover, for ${\rm Im}(z) \neq 0$, such divergences disappear
and $R$ tends to $\mathcal{O}(1)$ values like in the case of normal hierarchy. 

Let us briefly make here a comparison with other seesaws, still implemented in a SU(5) context. As already discussed in the previous section, in what respect neutrino masses our model is not different from a model with a type I seesaw with only two RH neutrinos. This statement holds also for LFV, with the only difference given by the fact that in the 2RHN model the heavy neutrino masses can be hierarchical, while our  model, barring cancellations, predicts $M_T\simeq M_S$. As a consequence higher values for $R$ can be obtained~\cite{Guo:2006qa}. In the type I seesaw with three RH neutrinos, due to the larger number of parameters, even more freedom is allowed. On the contrary in the type II seesaw there is a direct relation between high-energy and low-energy neutrino parameters, so that the ratios of the branching ratios of LFV processes can be expressed in terms of neutrino masses and mixing angles. When this is embedded into a SU(5) GUT by adding a ${\bf 15}$+${\bf \bar{15}}$ representation~\cite{Rossi:2002zb}, $R$ varies between 400 and $\mathcal{O}(1)$ with increasing $\theta_{13}$~\cite{Joaquim:2009vp}. Notice that in that model the seesaw fields induce flavour violation in the hadronic sector too, as in the model we are studying in this paper.

Let us now discuss the induced flavour violation in the hadronic sector. From Eq.~(\ref{eq:deltaRRd}), we see that off-diagonal entries
in the $m^2_{\tilde{d}^c}$ mass matrix are induced by the coupling  $\vec{y}_V$ of the down-quark SU(2) singlets with the {\bf 24} 
field $V$. Eqs.~(\ref{eq:yT}-\ref{eq:yV}) tell us that $\vec{y}_V$ cannot be unequivocally determined in terms of $\vec{y}_S$ and 
$\vec{y}_T$ and, therefore, in terms of neutrino parameters. However, Eqs.~(\ref{eq:yT}-\ref{eq:yV}) also show that $\vec{y}_V$
can be naturally expected to be of the same order of magnitude of the seesaw Yukawa couplings, as it clearly follows from the
SU(5) embedding of the model. In particular if $|\vec{y}_{1,2}|\ll |\vec{y}_{24}|$ in Eqs.~(\ref{eq:yT}-\ref{eq:yV}), then
$\vec{y}_V\simeq \vec{y}_T$. In our numerical analysis, we are going to make use of this last assumption\footnote{Another possible
approach to improve the predictivity of the model in the hadronic sector is considering the renormalizable version of the model
discussed in \cite{Perez:2007iw}, where $\vec{y}_V$ can be written as a combination of $\vec{y}_T$ and $\vec{y}_S$.
Apart from this point, that model gives the same phenomenology discussed here.}. 
Anyway, with a free choice of $\vec{y}_{1,2}$ and barring cancellations, we would still get:
\begin{equation}
 (\delta^d_{\rm RR})_{ij} \simeq  (\delta^e_{\rm LL})_{ij}  ~\frac{\overline{m}^2_{\tilde{L}}}{\overline{m}^2_{\tilde{d}^c}}\,.
\end{equation}
Thus, comparing this expression with Eq.~(\ref{eq:dLLestimate}), we find 
that the typical order of magnitude of the hadronic mass insertion is:
\begin{equation}
\label{eq:dRRestimate}
(\delta^d_{\rm RR})_{ij} \simeq \mathcal{O}(10^{-5}\div 10^{-4}) \quad \textrm{for}\  M_{V} \simeq 10^{13}~{\rm GeV}\,,  
\end{equation}
with $(\delta^d_{\rm RR})_{ij}$ becoming maximal for $m_0\gg M_{1/2}$, 
when $\overline{m}^2_{\tilde{d}^c}\simeq \overline{m}^2_{\tilde{L}}\simeq m^2_0$.
Moreover, one has to take into account that, like in the MSSM, the RGE for $m^2_{\tilde{Q}}$ 
generates also small $(\delta^d_{\rm LL})_{ij}$ proportional
to CKM elements, $(\delta^d_{\rm LL})_{ij}\propto V_{ti}^*V_{tj}$.
Taking into account this further effect, the most stringent bounds, which come from the Kaon system, 
are $|(\delta^d_{\rm RR})_{12}|\simeq \mathcal{O}(10^{-3})$ \cite{Altmannshofer:2009ne}. Comparing this value with
Eq.~(\ref{eq:dRRestimate}), we see that the model does not typically predict large deviations from the constrained MSSM (CMSSM) 
predictions in the hadronic sector and therefore it is safe from hadronic FCNC constraints. 
However, Eq.~(\ref{eq:dRRestimate}) provides a quite
rough estimate and it is worth to study in more detail some hadronic observables, for which experiments have recently showed 
possible tensions with the SM (and the CMSSM) predictions. In the next section, in particular, we are going to comment about the 
impact of the new flavour mixing sources of the model on the Kaon CP-violating parameter $\epsilon_K$ and on the time-dependent CP asymmetry, $S_{\psi\phi}$, in the decay $B_s\to J/\psi\phi$.

\section{Results}\label{Sec:4}

As mentioned above, in order to outline the 
effects induced by the RG running between the GUT scale and the mass scale
of the {\bf 24} fields, we consider universal boundary conditions, namely: a universal scalar mass $m_0$, a common gaugino mass $M_{1/2}$ and trilinear terms $A_f=A_0 \,Y_f$.

In order to compute the SUSY spectrum,
we numerically solve the full 1-loop RGEs of the model (see the Appendix)
down to the seesaw scale, $M_S = M_T \equiv M_{I}$, at which the {\bf 24} fields decouple.
Then, we run the MSSM RGEs down to the SUSY scale 
$m_{\rm SUSY} \equiv \sqrt{m_{\tilde{t}_1} m_{\tilde{t}_2}}$.
For each point of the parameter space, we impose 
the following requirements: (i) successful EWSB and absence of tachyonic particles;
(ii) limits on SUSY masses from direct searches; (iii) neutral LSP.
Then, we compute the leptonic processes by means of a full calculation
in the mass eigenstate basis \cite{lepton-processes}, 
the hadronic processes by means of the mass-insertion 
approximation formulae in \cite{MIs},
the LSP relic density using \texttt{DarkSUSY} \cite{darksusy} and
the ${\rm BR}(B\to X_s \gamma)$ using \texttt{SusyBSG} \cite{susybsg}.
We require that the resulting ${\rm BR}(B\to X_s \gamma)$ do not deviate from 
the experimental value \cite{HFAG} in more than 3$\sigma$. 

Let us first try to extract information about the seesaw scale and the other 
seesaw parameters, focusing on the stringent bounds LFV can impose on them. 
In order to do that, we can start varying all the parameters in large ranges, but 
we clearly need a criterion for defining the SUSY spectrum we want to concentrate on
(all effects would be negligibly small, if we considered slepton masses of several TeV). 
Therefore, we will mostly concentrate on parameter regions giving sizable SUSY contributions 
to the anomalous magnetic moment of the muon, $(g-2)_\mu$, and, later on, also on regions which
provide a dark matter relic density within the WMAP constraints. 

For simplicity, in the numerical analysis we have neglected possible $\mathcal{O}(1)$ mass-splittings among the fields in the {\bf 24}, inducing threshold corrections to the gauge coupling running, which would modify the MSSM gauge coupling unification. 
In particular, the 1-loop prediction for the value of the strong coupling at $M_Z$ would become:
\begin{equation}
 \frac{1}{\alpha_3(M_Z)}-\frac{1}{\alpha^0_3(M_Z)}= \frac{3}{2\pi}\left[\ln\frac{M_T}{M_O}+\frac{1}{7}\ln\frac{M_T}{M_V}\right]\,,
\end{equation}
where $\alpha^0_3(M_Z)$ is the 1-loop MSSM prediction. According to the above formula, the consistency with the measured value
for $\alpha_3(M_Z)$ could be slightly worsened or improved. We notice that such modification of the running of the gauge couplings would 
have anyway a negligible impact on the running of the other parameters. 
Moreover, the possible thresholds would affect the running of the soft masses, entering in the expressions of the flavour violating parameters, Eqs.~(\ref{eq:deltaLLe}, \ref{eq:deltaRRd}), only logarithmically, so that they would have a small effect on the observables we are going to study.
\begin{figure}[t!]
\includegraphics[width=0.6\linewidth, angle=-90]{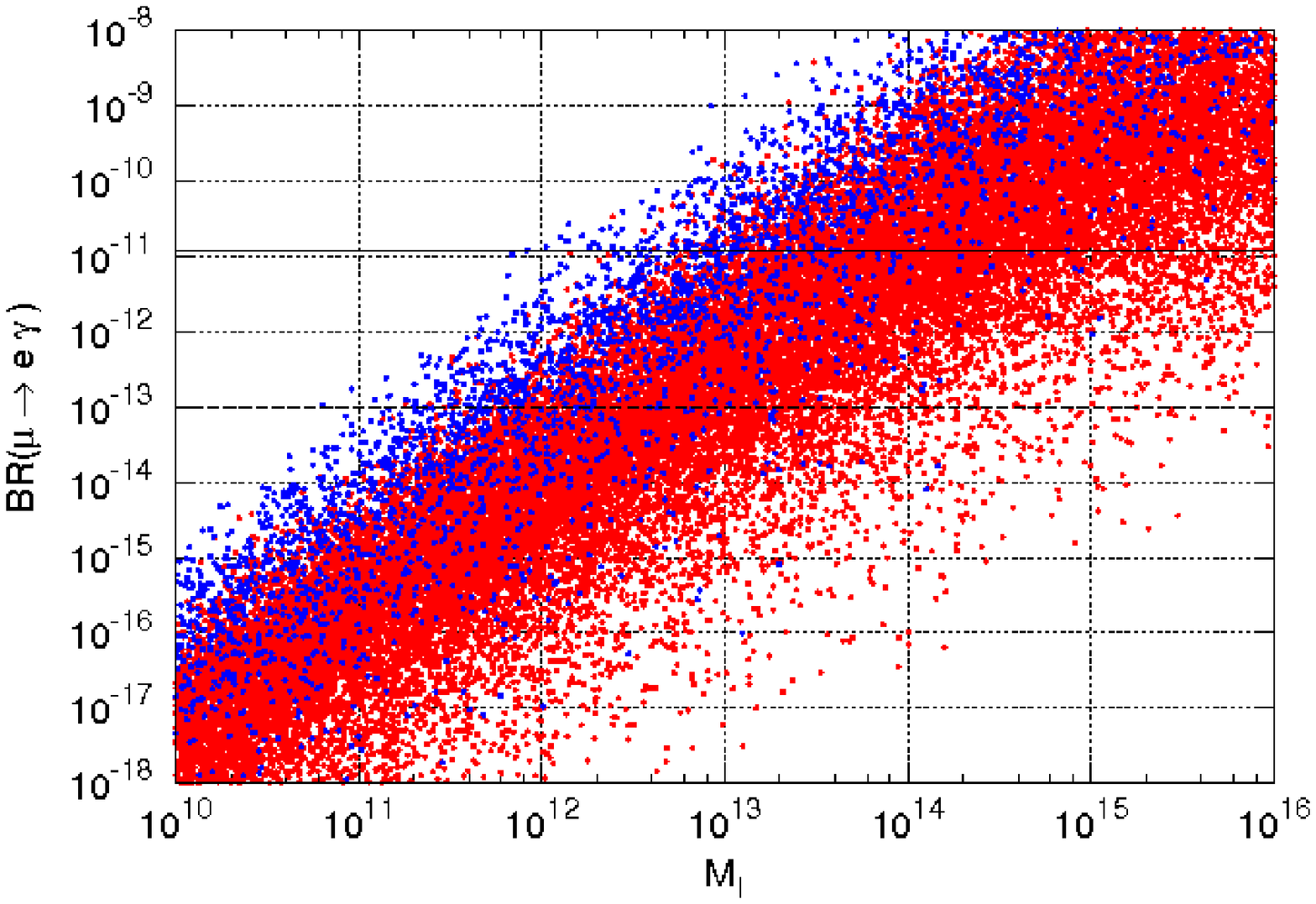}
\includegraphics[width=0.6\linewidth, angle=-90]{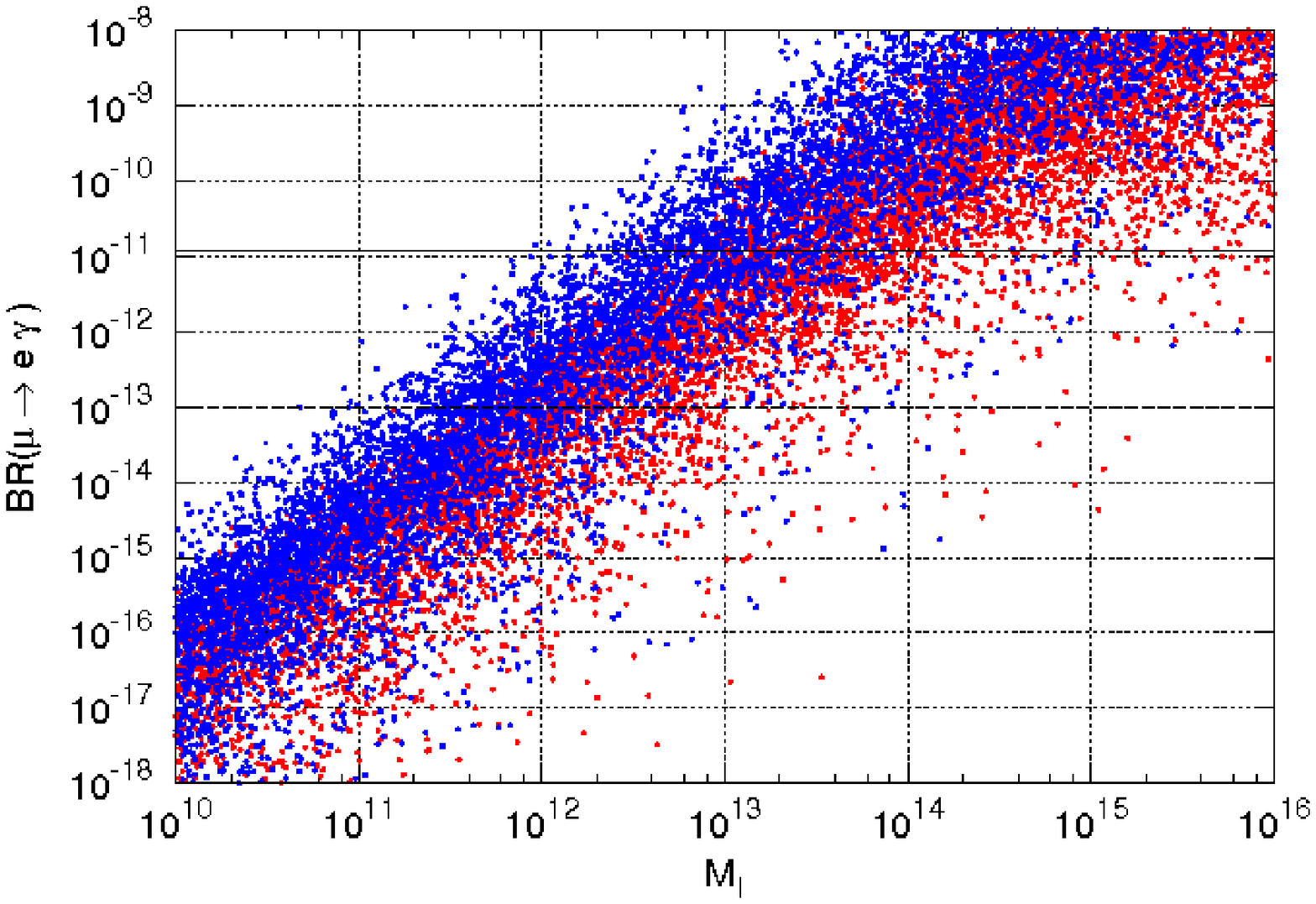}
\caption{BR$(\mu\to e\gamma)$ vs. the seesaw scale $M_I$ (in GeV) for $\tan\beta=10$ (top) and $\tan\beta=40$ (bottom)
and a wide scan of the parameters (see the text for details). Blue points provide $(g-2)^{\rm SUSY}_\mu/2 > 10^{-9}$.}
\label{fig:megvsMT}
\end{figure}

\subsection{Lepton Flavour Violation}
In Fig.~\ref{fig:megvsMT}, we plot BR$(\mu\to e\gamma)$ as a function 
of the seesaw scale $M_S = M_T \equiv M_{I}$, in the case of normal neutrino hierarchy\footnote{We checked that inverted hierarchical
neutrinos do not provide significantly different predictions with respect to the normal hierarchy case. We will thus
concentrate on normal hierarchy from now on.},
for the following choice of the parameters:
$\tan\beta=10$ (top panel) and $\tan\beta=40$ (bottom panel), 
$0 < m_0 < 1$ TeV, $0 < M_{1/2} < 1.5$ TeV, $|A_0| \leq 3 m_0$. 
The neutrino parameters were also varied in the following ranges:
$10^{10}~\textrm{GeV} \leq  M_{I} \leq 10^{16}$~GeV, $10^{-3} \leq |U_{e3}| \leq 0.2$, $0 \leq z < 2 \pi$.
We took the parameter $z$ real, since the only effect of its imaginary part is to raise the seesaw Yukawas 
and so the $\mu\to e\gamma$ rate, as we commented in Sec.~\ref{Sec:2}. However,
BR$(\mu\to e\gamma)$ itself provides very stringent bounds on ${\rm Im}(z)$, as we will comment below.
We have also checked that all couplings remain perturbative up to the GUT scale. 
The blue (black) points give $a^{\rm SUSY}_\mu\equiv(g-2)^{\rm SUSY}_\mu/2 > 10^{-9}$, so lowering
the tension between theoretical prediction and experiments below the 2$\sigma$ level. 

For $\tan\beta=10$ (upper panel of Fig.~\ref{fig:megvsMT}), 
we see, besides the dependence 
BR$(\mu\to e\gamma)\sim M_I^2$, that the current experimental limit
BR$(\mu\to e\gamma)< 1.2 \times 10^{-11}$ \cite{MEGA}, 
constrains the seesaw scale to be $M_I \lesssim 10^{13}\div 10^{14}$~GeV
for the points favored by $(g-2)_\mu$, even if there are few points, for which 
the parameters conspire in lowering BR$(\mu\to e\gamma)$, that can evade such bound. 
Even if BR$(\mu\to e\gamma)$ is enhanced by increasing $\tan\beta$, 
we find the above limit on the seesaw scale also for $\tan\beta=40$ (lower panel of Fig.~\ref{fig:megvsMT}), 
since $a^{\rm SUSY}_\mu$ is increased by $\tan\beta$ as well. 
In both cases, the MEG experiment \cite{meg}, whose expected sensitivity is BR$(\mu\to e\gamma)\simeq 10^{-13}$,
will be able to test soon the region of the parameter space favored by $(g-2)_\mu$ down to 
$M_I \simeq 10^{12}\div 10^{13}$~GeV.
\begin{figure}[t]
\includegraphics[scale=0.7]{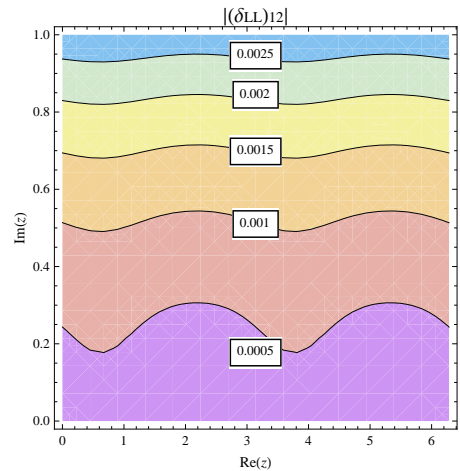}
\caption{Contour plot for the LFV parameter $|(\delta^e_{\rm LL})_{12}|$ in the Re($z$)-Im($z$) plane, for 
$m_0=M_{1/2}$, $A_0 = 0$, $M_I=10^{13}$ GeV and $U_{e3}=0$, $\Phi=0$.}
\label{fig:dLLbound}
\end{figure}
\begin{figure}[t]
\includegraphics[width=0.6\linewidth, angle=-90]{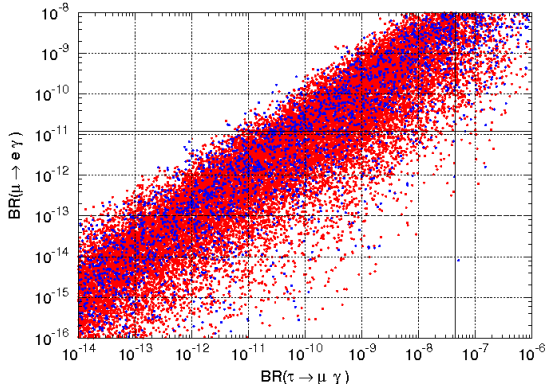}
\caption{BR$(\mu\to e\gamma)$ vs. BR$(\tau\to \mu\gamma)$, for $\tan\beta = 10$ and 
the same scan of the parameters of Fig.~\ref{fig:megvsMT}.}
\label{fig:meg-tmg}
\end{figure}

Let us now show how the present bound on BR$(\mu\to e\gamma)$ can severely constrain the parameter $z$. 
We have already argued above that Im($z$) cannot be too large, without having unnatural cancellations
in the neutrino mass matrix $m_\nu$ between the singlet and the triplet terms. Besides that,  
LFV bounds can directly constrain Im($z$), since the seesaw Yukawas simply grow by increasing it.
For convenience, let us express the BR$(\mu\to e\gamma)$ constraints in terms of bounds on the mass insertion
$(\delta^e_{\rm LL})_{12}$. From the same scan of the parameters presented above, we find that 
satisfying the present limit on $\mu\to e\gamma$ requires: 
\begin{equation}
|(\delta^e_{\rm LL})_{12}| < (5\div6)\times 10^{-4}\,,
\label{eq:dLLbound}
\end{equation}
for the points lying in the blue (black) region. 
In Fig.~\ref{fig:dLLbound}, we show contours for $|(\delta^e_{\rm LL})_{12}|$ in the Re($z$)-Im($z$) plane, for 
$m_0=M_{1/2}$, $A_0 = 0$, $M_I=10^{13}$ GeV and $U_{e3}=0$, $\Phi=0$. 
We see that, indeed, $|(\delta^e_{\rm LL})_{12}|$ grows very fast with Im($z$).
As a consequence, the bound of Eq.~(\ref{eq:dLLbound}) constrains Im($z$) to values $\lesssim 0.3$ for the seesaw scale
at $10^{13}$ GeV.

Let us finally consider LFV in the $\mu$-$\tau$ sector as well. In Fig.~\ref{fig:meg-tmg}, we plot 
BR$(\mu\to e\gamma)$ vs. BR$(\tau\to \mu\gamma)$, for $\tan\beta = 10$ and 
the same variation of the other parameters of Fig.~\ref{fig:megvsMT}. 
We see that in this model, the strong bound on flavour transition in the $\mu$-$e$ sector already challenges
future $\tau\to\mu\gamma$ experiments quite strongly. In fact, the bulk of the points, for which 
BR$(\mu\to e\gamma)$ is less than the present bound, gives BR$(\tau\to \mu\gamma)\lesssim 10^{-9}$, which is
indeed below the expected sensitivity of the proposed Super Flavour Factory \cite{superF}. This is consistent with  
the estimate for $R = {\rm BR}(\tau\to \mu\gamma)/{\rm BR}(\mu\to e\gamma) \lesssim 100$, we provided in the previous section.
However, we see that there are some points for which parameters conspire to raise the value of ${\rm BR}(\tau\to \mu\gamma)$
at the level of $10^{-8}$, i.e. in the reach of the SuperB factory at KEK \cite{KEK}.
Nevertheless, a positive signal for ${\rm BR}(\tau\to \mu\gamma)$ would anyway disfavor the scenario under study.

\subsection{Hadronic observables}
\begin{figure}[t]
\includegraphics[width=0.6\linewidth, angle=-90]{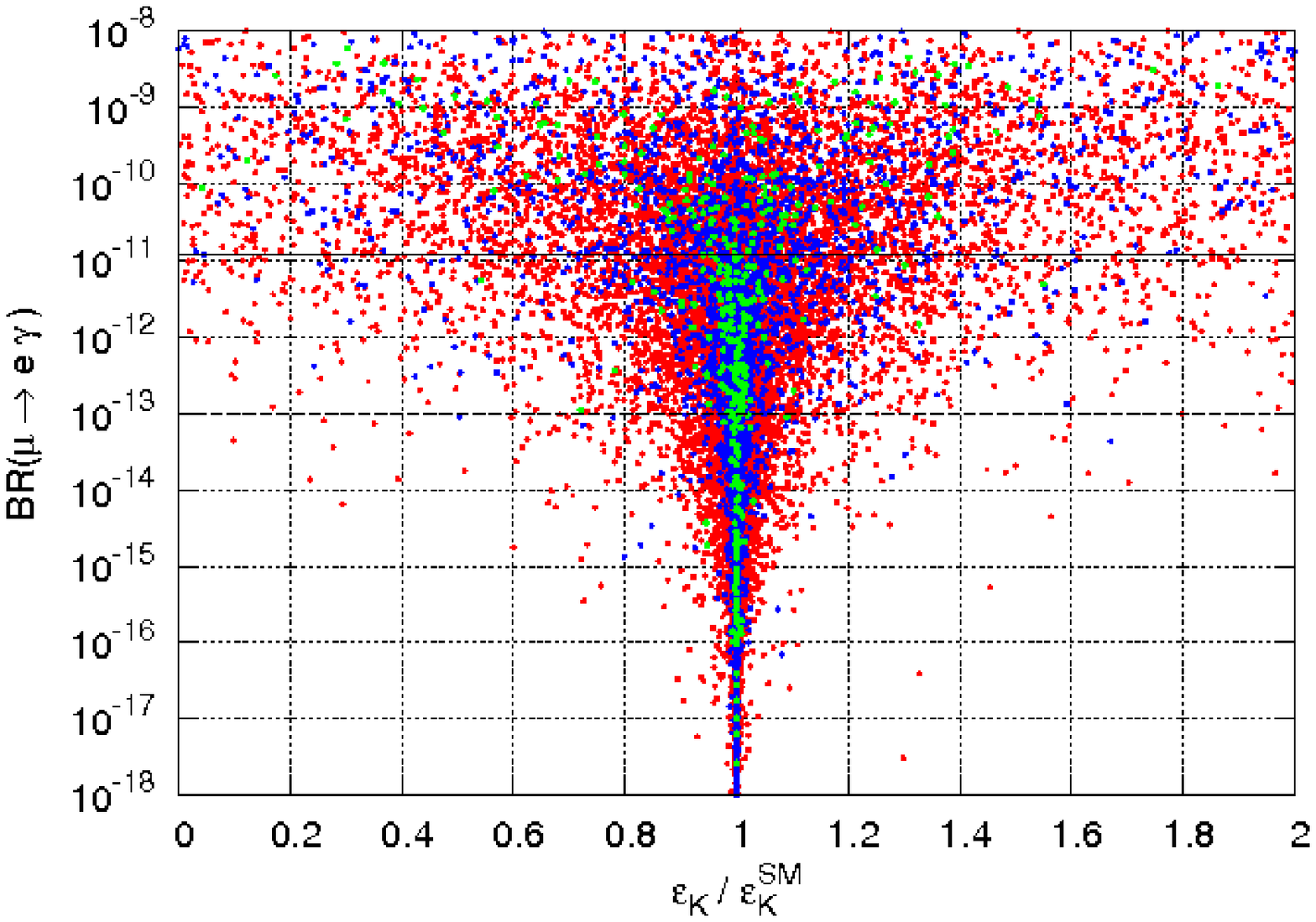}
\includegraphics[width=0.6\linewidth, angle=-90]{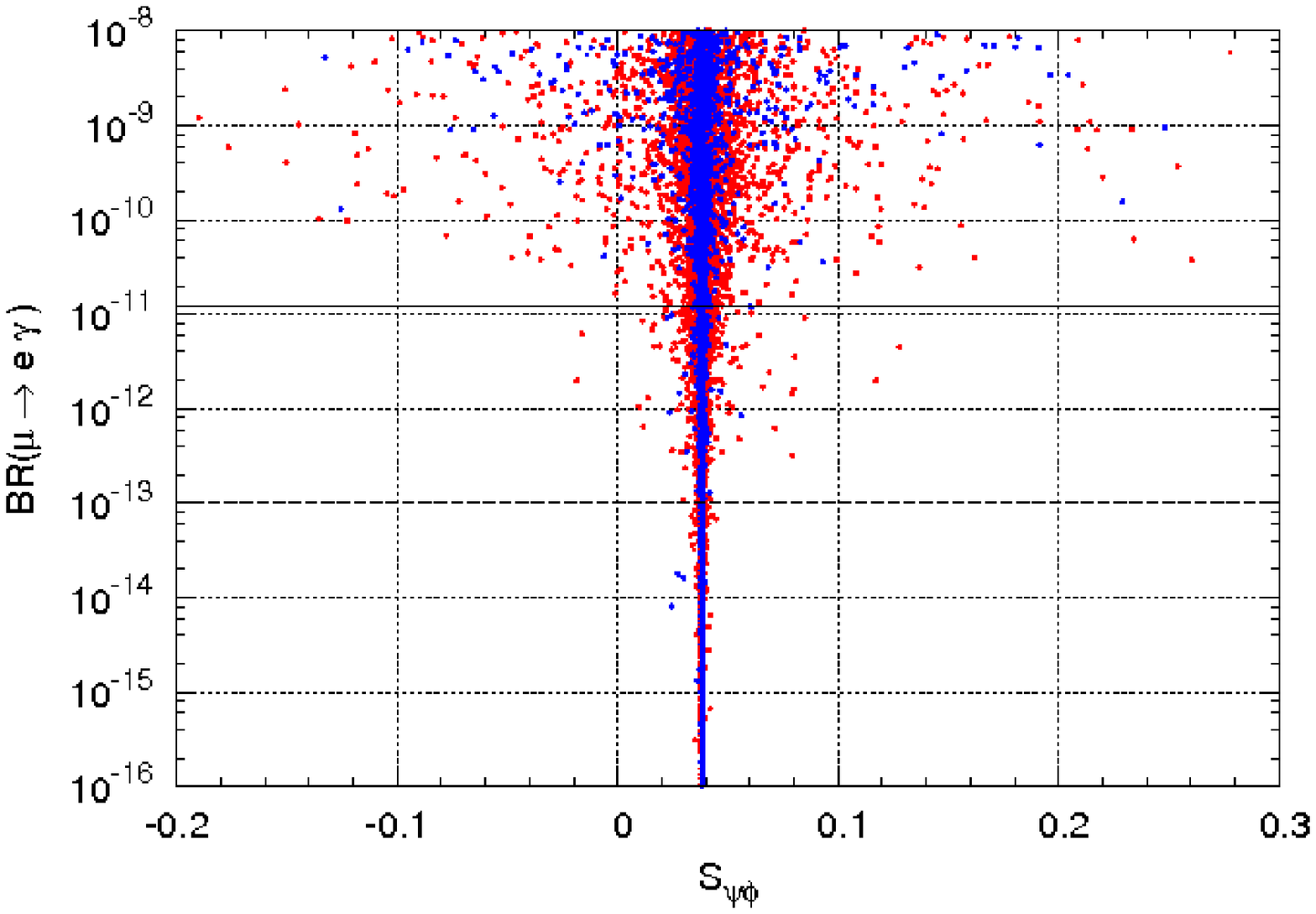}
\caption{Top: BR$(\mu\to e\gamma)$ vs. $\epsilon_K/\epsilon_K^{\rm SM}$ for
$10 \le \tan\beta \le 40$ and the same scan of parameters as in Fig.~\ref{fig:megvsMT}. Bottom:
BR$(\mu\to e\gamma)$ vs. $S_{\psi\phi}$.}
\label{fig:ek-Spsiphi}
\end{figure}

In this section, we are going to discuss the effects of the new source of flavour mixing in the down squark sector, Eq.~(\ref{eq:deltaRRd}), induced by the running between the GUT and the seesaw scales. In particular, it is interesting to check
if this is able to account for a large phase in the $B_s$ mixing, as suggested
by the Tevatron experiments CDF \cite{CDF} and D0 \cite{D01, D02}. Moreover, a positive new physics contribution to $\epsilon_K$ 
(around the 24\% of the SM contribution) \cite{buras-diego} is one of the possible ways for accommodating a recently reported tension
among different observables used to fit the unitarity triangle (see also \cite{Altmannshofer:2009ne, Soni}). 

As pointed out in section \ref{Sec:3}, hadronic flavour mixing cannot be directly related to the leptonic one. Nevertheless,
the off-diagonal entries of $m^2_{\tilde{d}^c}$ are naturally of the same order of magnitude as the leptonic ones.
For definitiveness, we take $\vec{y}_V=\vec{y}_T$ as input for the RGEs at the GUT scale, then we vary the phases of the resulting
low-energy mass insertions $(\delta^d_{\rm RR})_{ij}$ between 0 and 2$\pi$. 

Let us start to consider the possible effect of the generated $(\delta^d_{\rm RR})_{12}$ on the SUSY contribution to $\epsilon_K$.
In the top panel of Fig.~\ref{fig:ek-Spsiphi}, we plot BR$(\mu\to e\gamma)$ vs. $\epsilon_K/\epsilon_K^{\rm SM}$ for
$10 \le \tan\beta \le 40$ and the same variation of the parameters as in Fig.~\ref{fig:megvsMT}.  
As for the previous plots, the blue (black) points provide a sizable SUSY contribution to $(g-2)_\mu$, while the green (light grey)
points give a neutralino relic density not larger than the cold dark matter relic density measured by WMAP 
(see the next section for details). We can see that the present bound on BR$(\mu\to e\gamma)$ still allows for a sizable 
(up to 20$\div$30 \% of $\epsilon_K^{\rm SM}$) positive contribution to $\epsilon_K$. Furthermore 
we have BR$(\mu\to e\gamma) > 10^{-13}$ (so within the sensitivity of MEG) 
for most of the points which provide such a solution to the $\epsilon_K$ tension, which 
would be therefore strongly disfavored by a negative result of MEG. Let us notice, however, that the
parameter space points favored by WMAP cannot provide the desired increase of $\epsilon_K$. 
The reason is that these points are mainly concentrated in the coannihilation region where $M_{1/2} > m_0$ or even
$M_{1/2} \gg m_0$, as we are going to discuss in the next section, so that the flavour violating $(\delta^d_{\rm RR})_{ij}$ 
result suppressed by large squark masses.

In the bottom panel of Fig.~\ref{fig:ek-Spsiphi}, we plot $S_{\psi\phi}$ for the same scan of the parameters.
As we can see, the predicted value do not deviate too much from the small SM prediction 
$S^{\rm SM}_{\psi\phi}\simeq 0.036$. The reason is that, even if the phase of $(\delta^d_{\rm RR})_{23}$ can be 
large, $|(\delta^d_{\rm RR})_{23}|$ is numerically too small (cfr. for instance the estimate in Eq.~(\ref{eq:dRRestimate}))
to provide a sizable CP violation in the $B_s$ mixing and thus accounting for the di-muon anomaly reported in \cite{D02}.
If such new physics effects in $B_s$ mixing will be confirmed, the minimal version of the model we are discussing here should
be extended to include further sources of flavour violation in the hadronic sector.

\subsection{Neutralino relic density}
The presence of intermediate scale fields, which are charged under the SM gauge group, has a possible 
impact on the supersymmetric spectrum and, thus, on the parameter space regions, 
for which the relic density of the LSP (in our case a bino-like lightest neutralino as in the CMSSM) results
to be within the WMAP bounds \cite{wmap}. In this section, we are going to focus on the so-called $\tilde\tau$
coannihilation region \cite{coann}, since focus point \cite{focuspoint} and A-funnel \cite{polefunnel}
are not expected to be qualitatively different with respect to the CMSSM 
(even if they can be quantitatively modified, even significantly, within this model).

The effect we are going to discuss can be again traced back to the modification of the RG running of the parameters. 
In this case, however, this is not due to the new Yukawa interactions 
(since flavour bounds do not allow the couplings to be too large), but it is an effect of the 
modification of the running of the gauge couplings (and the gaugino masses) above the scale 
of the {\bf 24}. 
In fact, even if the fields in the {\bf 24} do not spoil (at least at 1-loop) 
the successful gauge coupling unification of the MSSM, 
the running gets ``stronger'': above $M_T\simeq M_V \simeq M_O$ the 1-loop 
$\beta$-function coefficients gets indeed modified as follows:
\begin{equation}
 b_i = b_i^{\rm MSSM} + b_i^{24} = (33/5, 1, -3) + (5,5,5)\, ,
\end{equation}
and the running of the gauge couplings is considerably deflected. 
As a consequence, even if the couplings unify at the usual MSSM GUT scale, $M_{\rm GUT}\sim 10^{16}$ GeV, 
the value of the unified coupling $\alpha_U$ gets larger than in the CMSSM.

Clearly, an analogous effect happens to the gaugino masses, so that they reach values at $M_I$, which
can be considerably smaller than the unified value $M_{1/2}$. This could be thought as a simple rescaling
of $M_{1/2}$ (since clearly the low-energy gaugino masses will be the same as in the CMSSM with a lower
value of $M_{1/2}$), if it did not affect the scalar masses as well. In fact, with the same values of the gaugino masses
at low energy, the scalar mass will feel a stronger gauge contribution to the running, 
through the gauge terms in the RGEs, $\sim \alpha_i M_i^2$, which are larger than in the CMSSM between the GUT scale and $M_I$.  
The consequence in the low-energy SUSY spectrum is that the scalar masses will result relatively larger, with respect to the gaugino masses,
than in the MSSM. 

Qualitatively, the above described effect is clearly common to all models that have fields charged under the SM gauge group at some
 intermediate scale, and it was, for instance, observed in the context of an $SO(10)$ type-II seesaw model in \cite{Calibbi:2009wk} 
and in a multi-scale flavour model in \cite{Calibbi:2008yj}.

Coming back to DM, having relatively heavier scalars could destabilize the ordinary regions of the parameter space that provide
a neutralino relic density within the WMAP bounds and for which quite precise relations among parameters are usually required. 
An example is the $\tilde{\tau}$ coannihilation region, where the correct relic density is achieved thanks to 
an efficient $\tilde{\tau}$-$\tilde{\chi}^0_1$ coannihilation, which requires $m_{\tilde{\tau}_1} \simeq m_{\tilde{\chi}^0_1}$. 
As we are going to see, such region is strongly modified in our case, as an effect of the relatively heavier $\tilde{\tau}_1$
resulting from the strong gauge running below the GUT scale. 
\begin{figure}[t]
\includegraphics[width=0.7\linewidth, angle=-90]{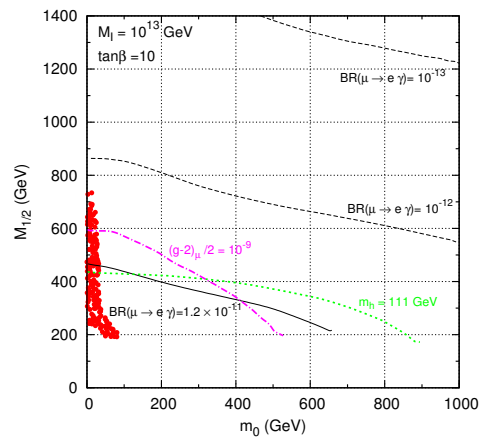}
\caption{Coannihilation region and ${\rm BR}(\mu\to e\gamma)$ contours in the ($m_0$, $M_{1/2}$) plane, for $\tan\beta=10$, $M_I = 10^{13}$ GeV.}
\label{fig:DMps}
\end{figure}

What can happen to the coannihilation region is depicted in Fig.~\ref{fig:DMps}, where we show the $m_0$-$M_{1/2}$ plane
for $M_I = 10^{13}$ GeV, $\tan\beta =10$, $A_0 = 0$. 
The neutrino parameters (not relevant for DM) were taken to be $U_{e3}=0$, $z = 1$.
The region marked with red (grey) points gives $\Omega_{\rm DM} h^2 \leq 0.13$. 
We can see that the CMSSM region where  $m_{\tilde{\tau}_1} < m_{\tilde{\chi}^0_1}$, along which usually the coannihilation strip runs,
has disappeared as a consequence of the effect described above\footnote{This opens up the possibility of having $m_0 =0$, 
i.e. vanishing scalar masses at high-energy (then generated
through the running driven by the gaugino masses), such as in \cite{Calibbi:2009wk, Calibbi:2008yj}. This possibility
has been recently addressed in \cite{ellis-olive}.}.
Coannihilation is still possible, since very low values of $m_0$ still gives $m_{\tilde{\tau}_1} \simeq m_{\tilde{\chi}^0_1}$,
but, interestingly, such region is bounded from above: this means that this particular set-up of the parameters 
predicts an upper bound on the DM mass, in this case $m_{\tilde{\chi}^0_1}\lesssim 240$ GeV, as we can see from the figure
taking into account that, for $M_I = 10^{13}$ GeV, the bino mass is approximately $M_1 \simeq 0.31\,M_{1/2}$.
A similar effect, providing an upper bound on $m_{\tilde{\chi}^0_1}$, was found in \cite{Calibbi:2007bk}, 
again as a consequence of the modification of the gauge contribution 
to the running of the scalar masses\footnote{See also \cite{Drees:2008tc, Carquin:2008gv, Gomez:2010ga}.} 
(in that case an additional SU(5) running of the parameters above the GUT scale was taken into account).
\begin{figure}[t]
\includegraphics[width=0.7\linewidth, angle=-90]{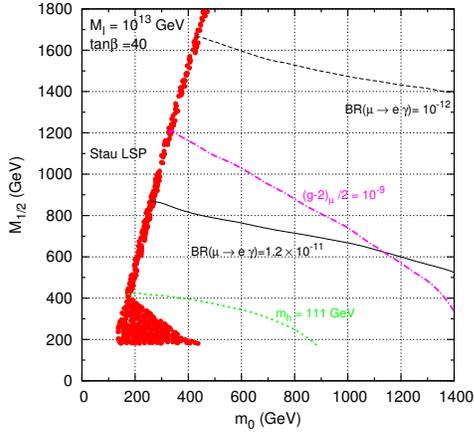}
\caption{Coannihilation region and ${\rm BR}(\mu\to e\gamma)$ contours in the ($m_0$, $M_{1/2}$) plane, 
for $\tan\beta=40$, $M_I = 10^{13}$ GeV.}
\label{fig:DMps40}
\end{figure}

In Fig.~\ref{fig:DMps}, we have also plotted contours for ${\rm BR}(\mu\to e\gamma)$, the LEP limit on the Higgs mass
(taking into account 3 GeV of theoretical error), as well as the region which provides $(g-2)^{\rm SUSY}_\mu/2 > 10^{-9}$
(below the magenta dot-dashed line). We can see that the DM region is already partially excluded by the present limits
on $\mu\to e\gamma$ and on the Higgs mass. The rest of the coannihilation region, which is, at least in part, consistent 
with a sizable $(g-2)^{\rm SUSY}$, will be fully tested very soon by MEG, since it gives ${\rm BR}(\mu\to e\gamma)> 10^{-12}$.
The prediction for the $\mu\to e\gamma$ rate clearly depends on the parameter $z$, which we have here taken $z=1$. 
Nevertheless, we checked that, varying the value of $z$, still ${\rm BR}(\mu\to e\gamma)$ is predicted 
in the reach of the MEG experiment (i.e. $\gtrsim 10^{-13}$) in the coannihilation region, 
apart from few points where the combination of parameters happens to suppress the rate. 

Finally, it is important to stress how the effect described above and its possible impact on the coannihilation region 
are sensitive to variations of the parameters, especially $M_I$ and $\tan\beta$. 
The effect would be clearly decreased, and would eventually disappear, by increasing $M_I$, i.e. decreasing 
the length of the running and thus the value of $\alpha_U$,
and vice-versa would become stronger for lower values of $M_I$. A larger value of $\tan\beta$ would contribute as usual to decrease
the ${\tilde\tau}_1$ mass (by increasing the negative contributions $\propto y_\tau^2$ in the running and also the L-R mixing term in the
$\tilde\tau$ mass matrix). This can be seen in Fig.~\ref{fig:DMps40}, where the $\tan\beta =40$ case is shown. 
The parameter space is now qualitatively similar to the CMSSM case: the region where ${\tilde\tau}_1$ is the LSP has reappeared
and the coannihilation region is a strip along it. Notice the presence for low values of $m_0$ and $M_{1/2}$ of a sizable 
``bulk'' region (which is smaller but still present also for $\tan\beta=10$).
This region is, however, already excluded by several constraints, including the experimental limit on ${\rm BR}(\mu\to e\gamma)$. 

\section{Discussion and conclusions}\label{Sec:5}
In this paper we have considered a SUSY SU(5) model where neutrino masses are obtained via a mixed type I+III seesaw mechanism and we have studied its phenomenology assuming universal soft masses at the GUT scale. The main characteristic of the model is the presence of one massless neutrino. Then the high-energy seesaw parameters are less than in the three massive neutrinos case and therefore a higher degree of predictability is present. 
Moreover, the model represents a very economical way of accounting for neutrino masses in a GUT context, since the addition of
just one chiral superfield in the SU(5) adjoint representation is considered. 

Besides discussing the model and in particular the mechanism through which we obtain two neutrino masses, we have analyzed the following features:
\begin{itemize}
\item $\mu \to e \gamma$ and other LFV processes;
\item possible contributions to hadronic observables;
\item neutralino relic density. 
\end{itemize}

We have shown that we can have sizable contribution to the $\mu\to e\gamma$ rate, such that the current experimental limit 
constrains the seesaw scale to be $M_I \lesssim 10^{13}\div 10^{14}$~GeV, while MEG will be able to test the model down to scales of $M_I \lesssim 10^{12}\div 10^{13}$ GeV. We have also shown that the bounds on BR($\mu\to e\gamma$) put strong constraints on BR($\tau\to\mu\gamma$), making very unlikely to observe it in future experiments. 
Otherwise, a positive signal for the $\tau\to\mu\gamma$ decay would disfavor this model.

From the bound on BR($\mu\to e\gamma$), we have been able to put an upper bound on Im$(z)\lesssim 0.3$, for the seesaw scale $M_I=10^{13}$~GeV. Of course this bound is $M_I$-dependent, since a reduction in the scale would imply a decrease of the size of Yukawas and then larger values for Im($z$) would be allowed. However, as discussed in Sect.~\ref{Sec:2}, values of Im($z$) larger than 1 are unnatural since cancellations in the neutrino sector would be needed.

The contribution in the hadronic sector is given by the coupling of the down-quark singlets with the new fields $V$. Even if this cannot be directly related to the neutrino parameters, an order of magnitude estimate can be performed. We have shown that in this model the present bound on BR$(\mu\to e\gamma)$ still allows for a sizable (up to 20$\div$30 \% of $\epsilon_K^{\rm SM}$) positive contribution to $\epsilon_K$, which would help in accommodating a recently reported tension among different observables used to fit the unitarity triangle. On the other side, CP violation in the $B_s$ mixing turns out to be too small to be able to account for the di-muon anomaly reported 
by the D0 collaboration.

As for the neutralino relic density, we have focussed on the so-called $\tilde\tau$ coannihilation region. We have shown that the CMSSM region where  $m_{\tilde{\tau}_1} < m_{\tilde{\chi}^0_1}$, along which usually the coannihilation strip runs,
can disappear, so that the coannihilation region gets distorted. Interestingly, such region is bounded from above, which means that an upper bound on the DM mass can be derived: for the particular set-up of the parameters considered here we got $m_{\tilde{\chi}^0_1}\lesssim 240$~GeV. Moreover, the possibility of having $m_0=0$ (with efficient coannihilation as well) 
as high-energy boundary condition is now open. 

In this paper we have not addressed other issues such as proton decay and leptogenesis. As for proton decay, this model does not improve the situation with respect to the standard case, so one has to rely on standard mechanisms to suppress the proton decay rate. For instance, in the context of the missing partner mechanism~\cite{DTsplit} for solving the doublet-triplet splitting problem, it is possible 
to build models in which the proton decay rate is sufficiently suppressed~\cite{AFM}. An extended Higgs sector is required in that case. 
In principle, our model could be embedded in such extended SU(5) framework. For a review about proton stability, including also a section
about SUSY SU(5) models, where further possibilities for suppressing the proton decay are discussed, we refer to~\cite{Nath:2006ut}.

For what concerns leptogenesis, we argue that it can be realized in this model through the decay of the triplet or the singlet or both. However, since their exact masses are not determined from phenomenological constraints (contrary to the non-SUSY case addressed in Ref.~\cite{Blanchet:2008cj}), it is not clear who is the responsible for leptogenesis: actually a combined action of the two could be possible, since their masses are of the same order of magnitude. To derive a definite conclusion as well as bounds on the parameters, a dedicated study would then be needed, which is beyond the scope of this work.
\\

%
\noindent \textit{Acknowledgments:} 
We are grateful to Michele Frigerio and Paride Paradisi for useful discussions. 
We also acknowledge the hospitality and partial support of the Galileo Galilei Institute for Theoretical
Physics (GGI), Firenze, where part of this work was carried out. 
%
%

\vspace{1cm}

\appendix 
\section{Renormalization group equations}\label{Sec:rge}
We present here the complete RGEs for this model. In the case of the MSSM parameters, 
we only explicitly write the new contributions to the 1-loop $\beta$-functions, 
according to the definition: 
\begin{equation}
16\pi^2\frac{d}{dt} X =\beta^{\rm MSSM, 1}_{X} + \beta^{{\bf 24},1}_{X} ,
\end{equation}
where $X$ can represent either Yukawa couplings, A-terms or soft masses.
Clearly, all $\beta^{{\bf 24},1}_{X}$ vanishes below the scale of the involved {\bf 24} fields.
The corresponding $\beta^{\rm MSSM, 1}_{X}$ can be found, for instance, in \cite{Martin:1993zk}.
For the new non-MSSM parameters, we provide the complete 1-loop $\beta$-functions,
still denoted as $\beta^{{\bf 24},1}_{X}$.

\begin{widetext}
%
%
We first write the $\beta^{{\bf 24},1}_{X}$ for the Yukawa couplings:
%
\begin{align}
\label{eq:betaYu}
\left(\beta^{{\bf 24},1}_{Y_u}\right)_{ij} &=
\left(\frac{3}{2} |\vec{y}_T|^2 + |\vec{y}_S|^2+ 3 |\vec{y}_V|^2 \right) (Y_{u})_{ij} \, ,\\
%
\label{eq:betaYd}
\left(\beta^{{\bf 24},1}_{Y_d}\right)_{ij} &= 2 \,y_{V i}(\vec{y}_V^{\, \dagger} Y_d)_j  \, ,\\
%
\label{eq:betaYe}
\left(\beta^{{\bf 24},1}_{Y_e}\right)_{ij} &=
\frac{3}{2} (Y_e \vec{y}_T^{\, *})_i \,y^T_{T\,j}+ (Y_e \vec{y}_S^{\, *})_i \,y^T_{S\, j} \, ,\\
%
\label{eq:betaYS}
\left(\beta^{{\bf 24},1}_{y_S}\right)_i &=
\Big(-\frac{3}{5} g_1^2 - 3 g_2^2 
+ 3 {\rm Tr}(Y_u Y_u^{\dagger})+ 
 4\left|\vec{y_S}\right|^2
+\frac{3}{2} \left|\vec{y_T}\right|^2+
3 \left|\vec{y_V}\right|^2\Big) y_{S i} +
\frac{3}{2} \left(\vec{y_T^*}\cdot\vec{y_S}\right) y_{T i}  
+ (Y_e^T Y_e^* \vec{y}_S)_i   \, ,\\
%
\label{eq:betaYT}
\left(\beta^{{\bf 24},1}_{y_T}\right)_i &=
\Big(-\frac{3}{5} g_1^2 - 7 g_2^2 
+ 3 {\rm Tr}(Y_u Y_u^{\dagger})+ 
 \left|\vec{y_S}\right|^2
+4 \left|\vec{y_T}\right|^2
+3\left|\vec{y_V}\right|^2\Big) y_{T i} +
 \left(\vec{y_S^*}\cdot\vec{y_T}\right) y_{S i}
+ (Y_e^t Y_e^* \vec{y}_T)_i  \, ,\\
%
\label{eq:betaYV}
\left(\beta^{{\bf 24},1}_{y_V}\right)_i &=
\Big(-\frac{19}{15} g_1^2 - 3 g_2^2 -\frac{16}{3} g_3^2
+ 3 {\rm Tr}(Y_u Y_u^{\dagger})+ 
 \left|\vec{y_S}\right|^2
+\frac{3}{2} \left|\vec{y_T}\right|^2
+6 \left|\vec{y_V}\right|^2\Big) y_{V i}
+2 (Y_d Y_d^\dagger \vec{y}_V)_i\, .
\end{align}
%

The 1-loop $\beta$-functions for the soft scalar masses read:
%
%
\begin{align}
\label{eq:betamLa}
\left(\beta^{{\bf 24},1}_{m_{\tilde L}^2}\right)_{ij}=&~
\frac{3}{2}\Big(
y^*_{Ti}\, (\vec{y}_T^{\,T}\, m_{\tilde L}^2)_j\, +\, (m_{\tilde L}^2\, \vec{y}_T^{\, *})_i\, y^T_{Tj}+
 2\, y^*_{Ti}\,y^T_{Tj}\,(m_{H_u}^2+m_{\tilde T}^2)+2\, A^*_{Ti}\,A^T_{Tj}\Big)+\nonumber\\
&~ \Big(
y^*_{Si}\, (\vec{y}_S^{\,T}\, m_{\tilde L}^2)_j\, +\, (m_{\tilde L}^2\, \vec{y}_S^{\,*})_i\, y^T_{Sj}+
 2\, y^*_{Si}\,y^T_{Sj}\,(m_{H_u}^2+m_{\tilde S}^2)+2\, A^*_{Si}\,A^T_{Sj}\Big)\, ,
\\
\label{eq:betamda}
\left(\beta^{{\bf 24},1}_{m_{\tilde{d}^c}^2}\right)_{ij} =& ~
2\Big(
y^*_{Vi}\, (\vec{y}_V^{\,T}\, m_{\tilde{d}^c}^2)_j\, +\, (m_{\tilde{d}^c}^2\, \vec{y}_V^{\,*})_i\, y^T_{Vj}+
 2\, y^*_{Vi}\,y^T_{Vj}\,(m_{H_u}^2+m_{\tilde S}^2)+2\, A^*_{Vi}\,A^T_{Vj}\Big)\, , \\
%
\label{eq:betamHu}
\beta^{{\bf 24},1}_{m_{H_u}^2}=& ~
2\Big(
\frac{3}{2} \vec{y}^{\,\dagger}_{T} m_{\tilde L}^2 \vec{y}_T + 
\vec{y}^{\,\dagger}_{S} m_{\tilde L}^2 \vec{y}_S + 
3 \vec{y}^{\,\dagger}_{V} m_{{\tilde d}^c}^2 \vec{y}_V + 
 m_{H_u}^2(\frac{3}{2} |\vec{y_T}|^2+|\vec{y_S}|^2+3|\vec{y_V}|^2 ) + \nonumber\\ 
& ~\frac{3}{2} m_{T}^2|\vec{y_T}|^2+\,m_{S}^2|\vec{y_S}|^2+\,3  m_{V}^2|\vec{y_V}|^2+ 
 \frac{3}{2} |\vec{A_{T}}|^2+|\vec{A_{S}}|^2+3|\vec{A_{V}}|^2\Big) \,, \\
%
\label{eq:betamS}
\beta^{{\bf 24},1}_{m_S^2}=& ~
4\Big(
\vec{y}^{\,\dagger}_{S} m_{\tilde L}^2 \vec{y}_S +\, 
(m_{H_u}^2+m_{S}^2)|\vec{y_S}|^2+\, |\vec{A_{S}}|^2\Big)  \,, \\
%
\label{eq:betamT}
\beta^{{\bf 24},1}_{m_T^2} =& ~
2\Big(
\vec{y}^{\,\dagger}_{T} m_{\tilde L}^2 \vec{y}_T +\, 
(m_{H_u}^2+m_{T}^2)|\vec{y_T}|^2+\, |\vec{A_{T}}|^2\Big)-
16 M_2^2 g_2^2  \,, \\
%
\label{eq:betamV}
\beta^{{\bf 24},1}_{m_V^2} =& ~
2\Big(
\vec{y}^{\,\dagger}_{V} m_{\tilde L}^2 \vec{y}_V +\, 
(m_{H_u}^2+m_{V}^2)|\vec{y_V}|^2+\, |\vec{A_{V}}|^2\Big)-
\frac{10}{3} M_1^2 g_1^2- 6 M_2^2 g_2^2 -\frac{32}{3} M_3^2 g_3^2- g_1^2 S  \,, 
%
\end{align}
while $(\beta^{{\bf 24},1}_{m_{\tilde Q}^2})_{ij} =  (\beta^{{\bf 24},1}_{m_{{\tilde u}^c}^2})_{ij} = 
(\beta^{{\bf 24},1}_{m_{{\tilde e}^c}^2})_{ij} = \beta^{{\bf 24},1}_{m_{H_d}^2} = 0 $.
The hypercharge D-term contribution $S$ is given by:
\begin{equation}
{S}\ =\ m_{H_u}^2 - m_{H_d}^2 + {\rm Tr}(-m_{\tilde L}^2+m_{{\tilde e}^c}^2-2m_{{\tilde u}^c}^2+m_{{\tilde d}^c}^2+
m_{\tilde Q}^2) + 5 (m_V^2 - m_{\bar{V}}^2) \,.
\end{equation}

Let us finally write the $\beta$-functions for the trilinear terms:
%
%
%
\begin{align}
\label{eq:betaAu}
\left(\beta^{{\bf 24},1}_{A_u}\right)_{ij}=&~
\Big(\frac{3}{2}|\vec{y}_T|^2+|\vec{y}_S|^2+3|\vec{y}_V|^2\Big) (A_u)_{ij} +
\Big(3 \vec{y}_T\cdot\vec{A}_T+2 \vec{y}_S\cdot\vec{A}_S
+6\vec{y}_V\cdot\vec{A}_V\Big) (Y_u)_{ij} \,,\\
%
\label{eq:betaAd}
\left(\beta^{{\bf 24},1}_{A_d}\right)_{ij}=&~
2 y_{Vi} (\vec{y}_V^{\,\dag} A_d)_j +4 A_{Vi} (\vec{y}_V^{\,\dag} Y_d)_j\,,\\
%
\label{eq:betaAe}
\left(\beta^{{\bf 24},1}_{A_e}\right)_{ij}=&~
\frac{3}{2}(A_e \vec{y}_T^{\,*})_i y_{Tj}^T+(A_e \vec{y}_S^{\,*})_i y_{Sj}^T +
3(Y_e \vec{y}_T^{\,*})_i A_{Tj}^T+ 2 (Y_e \vec{y}_S^{\,*})_i A_{Sj}^T\,,
%
\end{align}
\begin{align}
%
\label{eq:betaAS}
\left(\beta^{{\bf 24},1}_{A_S}\right)_i =&~
 -\frac{3}{5}(A_{Si}-2 M_1 y_{Si}) g_1^2 - 3 (A_{Si}-2 M_2 y_{Si})g_2^2+ 
 \Big( 3{\rm Tr}(Y_u Y_u^{\dagger})+ \frac{3}{2}|\vec{y}_T|^2
+ 5 |\vec{y}_S|^2+3 |\vec{y}_V|^2\Big) A_{Si}+ \nonumber \\
&~ \Big( 6{\rm Tr}(A_u Y_u^{\dagger})+3 \vec{y}_T^{\, *}\cdot\vec{A}_T
+7 \vec{y}_S^{\,*}\cdot\vec{A}_S+ 6\vec{y}_V^{\,*}\cdot\vec{A}_V\Big) y_{Si}+ \nonumber \\
 &~ (Y_e^T Y_e^* \vec{A}_S)_i + 2 (A_e^T Y_e^* \vec{y}_S)_i + \frac{3}{2}(\vec{y}_T^{\,*}\cdot \vec{y}_S)A_{Ti}+
 3 (\vec{y}_T^{\,*}\cdot \vec{A}_S)y_{Ti}\,, \\
%
\label{eq:betaAT}
 \left(\beta^{{\bf 24},1}_{A_T}\right)_i =& ~-\frac{3}{5}(A_{T i}-2 M_1 y_{T i}) g_1^2 - 7 (A_{Ti}-2 M_2 y_{Ti})g_2^2+ 
 \Big( 3{\rm Tr}(Y_u Y_u^{\dagger})+ \frac{5}{2}|\vec{y}_T|^2
+ |\vec{y}_S|^2+ 3 |\vec{y}_V|^2\Big) A_{T i}+ \nonumber \\
&~ \Big( 6{\rm Tr}(A_u Y_u^{\dagger})+\frac{13}{2} \vec{y}_T^{\, *}\cdot\vec{A}_T
+2 \vec{y}_S^{\,*}\cdot\vec{A}_S+ 6\vec{y}_V^{\,*}\cdot\vec{A}_V\Big) y_{Ti}+ \nonumber \\
 &~ (Y_e^T Y_e^* \vec{A}_T)_i + 2 (A_e^T Y_e^* \vec{y}_T)_i + (\vec{y}_S^{\,*}\cdot \vec{y}_T)A_{S i}+
 2 (\vec{y}_S^{\,*}\cdot \vec{A}_T)y_{Si}\,, \\
%
\label{eq:betaAV}
\left(\beta^{{\bf 24},1}_{A_V}\right)_i =&
 -\frac{19}{15}(A_{Vi}-2M_1 y_{Vi}) g_1^2 -
 3 (A_{Vi}-2M_2 y_{Vi})g_2^2
-\frac{16}{3}(A_{Vi}-2 M_1 y_{Vi}) g_3^2+ \nonumber \\
&~ \Big( 3{\rm Tr}(Y_u Y_u^{\dagger})+\frac{3}{2}|\vec{y}_T|^2
+ |\vec{y}_S|^2+ 8 |\vec{y}_V|^2\Big) A_{Vi}+ 
 \Big( 3{\rm Tr}(A_u Y_u^{\dagger})+ 3\vec{y}_T^{\,*}\cdot\vec{A}_T
+2\vec{y}_S^{\,*}\cdot\vec{A}_S+ 11\vec{y}_V^{\,*}\cdot\vec{A}_V\Big) y_{Vi}+ \nonumber \\
 &~  2 (Y_d Y_d^\dag \vec{A}_V)_i + 4 (A_d Y_d^\dag \vec{y}_V)_i\,.
\end{align}
\end{widetext}



\end{document}